\documentclass[10pt,conference]{IEEEtran}
\usepackage[utf8]{inputenc}
\usepackage[T1]{fontenc}
\usepackage{float}
\usepackage{amsmath}
\usepackage{amsfonts}
\usepackage{mathtools}
\usepackage{amsfonts}
\IEEEoverridecommandlockouts
\usepackage{amssymb}
\usepackage{makeidx}
\usepackage{csquotes}
\usepackage{graphicx}
\usepackage{cite}
\usepackage{xcolor}
\usepackage{caption}
\usepackage{graphicx}
\usepackage{balance}
\usepackage[capitalise,nameinlink ,noabbrev]{cleveref}
\usepackage[algo2e,ruled, vlined, linesnumbered]{algorithm2e}
\usepackage{ifpdf}
\ifpdf
\else
  
\fi
\usepackage[margin=1in]{geometry}
\begin{document}
\title{ Enhancing Congestion Control to Improve User Experience in IoT Using LSTM Network}

\author{Atta Ur Rahman$^\dagger$, Bibi Saqia$^\dagger$, Wali Ullah Khan$^\ddagger$, Khaled Rabie$^\star$,\\ Mahmood Alam$^{\perp}$, Khairullah Khan$^\dagger$\\
$^\dagger$Department of Computer Science, University of Science and Technology, Bannu, Pakistan
\\
$^\ddagger$Interdisciplinary Centre for Security, Reliability and Trust, University of Luxembourg, Luxembourg\\
$^{\star}$Department of Engineering, Manchester Metropolitan University, Manchester, United Kingdom\\
$^{\perp}$School of Computer Science and Engineering, Central South University, Changsha Hunan, China\\
\{atta.rahman,saqiaktk,khairullah\}@ustb.edu.pk\\ waliullah.khan@uni.lu; k.rabie@mmu.ac.uk; 204708006@csu.edu.cn
}

\maketitle

\begin{abstract}
In the constantly developing realm of the Internet of Things (IoT), guaranteeing fast data transfer and a smooth user experience is critical. In IoT contexts with limited resources, congestion control is crucial for sustaining network performance. This study suggests a new strategy for improving congestion control by deploying Long Short-Term Memory (LSTM) networks. LSTMs are recurrent neural networks (RNN), that excel at capturing temporal relationships and patterns in data. IoT-specific data such as network traffic patterns, device interactions, and congestion occurrences are gathered and analyzed.  The gathered data is used to create and train an LSTM network architecture specific to the IoT environment. Then, the LSTM model's predictive skills are incorporated into the congestion control methods. This work intends to optimize congestion management methods using LSTM networks, which results in increased user satisfaction and dependable IoT connectivity. Utilizing metrics like throughput, latency, packet loss, and user satisfaction, the success of the suggested strategy is evaluated. Evaluation of performance includes rigorous testing and comparison to conventional congestion control methods. The findings illustrate the concrete advantages of LSTM-enhanced congestion control in IoT, highlighting its potential to reduce network congestion and improve the overall user experience.

\end{abstract}


\section{Introduction}
The fast expansion of the IoT has brought in an era of extraordinary connectedness and data-driven innovation \cite{lim2019survey}. IoT has impacted several industries, including healthcare, manufacturing, transportation, and smart cities \cite{umair2021impact}. This has allowed for a wide range of applications that improve productivity, convenience, and quality of life \cite{alghazzawi2021congestion}. IoT offers a standard framework for interconnecting various devices over the internet. The outcome is an exponential growth in the number of connected gadgets on the Internet. Internet traffic is thus becoming more congested \cite{verma2020iot}. Modern traffic patterns are changing as a result of expanding tendencies towards smart mobility via sensor-equipped devices \cite{makarfi2020toward}. The main need for wireless communication systems is lossless, continuous transmission at a high data rate \cite{rasheed2022lstm}. AI can revolutionize our communication and interaction styles, and it is a strong tool for supporting the next stage of IoT gadgets and networks. Backscatter Communication (BC) and AI may be combined to enable a variety of novel uses that were not feasible earlier, as well as new possibilities for communication that is both inexpensive and energy-efficient \cite{ahmed2023state}. The backscatter sensor tag (BST) sends data to IoT gadgets by reflecting and modifying the source's imposed signal. In \cite{ahmed2022backscatter}, a novel framework for optimization is presented that concurrently maximizes the overall power of each source under poor sequential interference cancellation decryption. 
One of the hardest jobs to do while trying to increase a network's quality of service (QoS) in the IoT is congestion control. This is primarily due to the fact that modern wireless networks have a massive number of connections \cite{aimtongkham2021enhanced}. The congestion situation can lead to increased energy resource usage in constrained devices in addition to having negative impacts on the network's overall efficiency in terms of throughput and latency \cite{akpakwu2020cacc}. For instance, because the majority of IoT devices run on batteries, considerations for energy consumption must be included in all layers of network architecture design, notably in congestion control \cite{betzler2016coap}.

The IoT is a network of billions of intelligent devices that are connected through various forms of communication \cite{lorincz2021comprehensive}. IoT devices produce tremendous amounts of data, but their processing speed, battery consumption, and buffer capacity are still restricted. Smart device limitations are a major factor in IoT network congestion, which impairs performance and results in data loss \cite{jain2022congestion}.
Congestion control in IoT plays a critical role in achieving service performance requirements because of the restricted quantity of resources in IoT, including network bandwidth, node processing capabilities, and server capacities \cite{huang2014modeling}.
It is crucial to provide effective data transfer and a flawless user experience in this changing environment. IoT networks create substantial issues because of their intrinsic complexity, which is characterized by a variety of devices, shifting network circumstances, and resource constraints \cite{mishra2018analysis}.

Transmission Control Protocol (TCP) congestion control is implemented utilizing grid topology networks in the micro IP (uIP) stack \cite{lim2020improving}. They conducted a preliminary test employing the Cooja network simulator, and the findings reveal that the original uIP TCP leads to a large number of retransmissions when a radio-duty cycling technique like ContikiMAC is applied. It takes a lot of network nodes and resource constraints to design congestion management algorithms for group communication \cite{mondal2019iot}. A disadvantage of this practice is its inability to regulate high levels of network congestion, which reduces the effectiveness of communications and ultimately leads to data loss \cite{suwannapong2021encoco}. 
In order to reduce the likelihood of connecting in the pathway from the pre-congested node, the back-pressure approach was employed in \cite{homaei2020enhanced} for congestion control. This method relies on the quality of the queue. In order to distribute network traffic along the best pathways and reduce congestion, they acquire the path selection by applying a fuzzy decision-making algorithm.
IoT uses wireless sensor networks (WSN) for a variety of purposes, including data transmission via the Internet and environmental sensing. WSN-based IoT experiences increased packet loss rates, longer delays, and reduced throughput as a result of the problem of excessive congestion \cite{kaur2022green}. When a system receives more data, machine learning (ML) enhances its powers and intelligence. A support vector machine (SVM)-based model was presented in the research \cite{ata2021adaptive} to examine traffic congestion in the context of a smart city. The suggested model includes an IoT-based ML-enabled road traffic congestion management system that notifies users when there is congestion at a particular location. Similarly, ML methods are utilized in the study \cite{kasthuribai2021optimized} to identify the congested data in IoT. They employed a support vector machine (SVM) based on the artificial flora algorithm (AF) to reduce congestion in WSN-based IoT.
In order to deliver packets to a destination concurrently, 5G uses IoT to operate in high-traffic networks with numerous sensors. The 5G network delivers an enormous amount of capacity, less latency, and very fast data transmission rates. Therefore, it uses the stream control transmission protocol (SCTP), however, the typical SCTP's congestion control measures have an adverse impact on overall performance \cite{najm2019machine}.

In IoT contexts, network congestion is one of the major problems \cite{guillen2021intelligent}. When the demand for network resources exceeds the capacity, congestion results, which can cause performance degradation, increased delay, and even data loss. Effective congestion control is required to maintain network stability, provide timely data transportation, and maintain customer satisfaction \cite{hussain2021comparative}. While traditional congestion control strategies are appropriate for regular networks, they may fail to accommodate the specific characteristics and needs of IoT. In an effort to adapt to the dynamic and varied traffic patterns displayed by IoT devices, they frequently adopted either rule-based or heuristic-driven approaches, however, these methods frequently failed to yield reliable results \cite{li2018qtcp}.\\
In order to overcome these difficulties, this study suggests a cutting-edge strategy that makes use of LSTM networks, which are created to recognize temporal connections and patterns in sequential data. This work attempts to revolutionize congestion management tactics in IoT settings by utilizing their predictive capabilities. The use of LSTM networks for congestion prediction and control has the potential to enhance user experiences, boost network performance, and expand the possibilities of IoT applications. The proposed approach considerably reduces retransmission and hence energy usage while boosting network performance in terms of both throughput and transmission delay.

\section{Proposed Framework}
The IoT links everything, and as the number of connected objects grows, network congestion grows as well. Additionally, it consists of a wide range of devices, each with unique requirements. As a result, it requires a congestion management strategy that can handle all of the needs of various devices \cite{gheisari2019cccla}. The proposed approach attempts to improve congestion management strategies in IoT settings by employing LSTM networks. The process comprises gathering data, designing and training an LSTM network, integrating congestion management, and evaluating performance, as shown in Figure \ref{fig:prpoosed}. The process is outlined in the following steps:

\begin{figure}[t]
    \centering
    \includegraphics[width=0.9\linewidth]{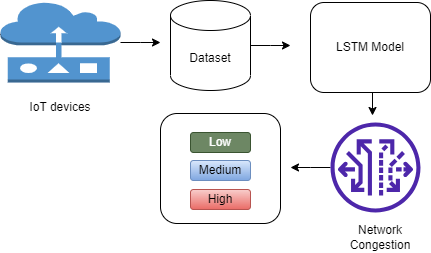}
    \caption{The steps involved in the proposed model}
    \label{fig:prpoosed}
\end{figure}

\subsection{Data Collection}
Gather IoT-specific data, such as network traffic patterns, device interactions, and examples of congestion occurrences. Data regarding network traffic comprises characteristics such as packet loss, delay, throughput, and the total number of active devices. This data is acquired from routers, switches, and IoT gateways. Keep track of connections between IoT devices, such as communication frequency, data transfer rates, and different types of messages transmitted. Determine the occurrences that point to congestion, such as a rise in packet loss, an expansion in latency, or a decrease in data transfer rates. To prepare the data for LSTM processing, normalize numerical features, encode categorical variables, and organize the data into sequences.

\subsection{Model Designing}

\begin{figure}[t]
    \centering
    \includegraphics[width=\linewidth]{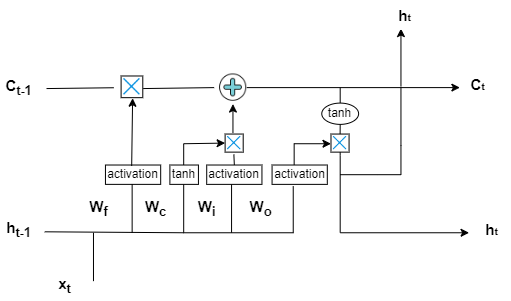}
    \caption{The LSTM model used for congestion control}
    \label{fig:lstm-model}
\end{figure}

Developing and training an LSTM network for congestion management includes building the architecture, specifying the input sequences, describing the layers, and training the model with gathered and preprocessed data. 
The LSTM architecture is made to forecast instances of congestion from previous sequences \(X\) of network traffic and congestion patterns. Each input sequence \(X_i\) is represented by a matrix of dimensions \(T \times F\), where \(T\) stands for the number of time steps and \(F\) for the number of features. We encode the target labels using a one-hot encoding scheme.\\ 
The LSTM network repeatedly analyzes the input sequence using its cells, modifying the hidden state $h_t$ and cell state $C_t$ at each time interval $t$. For each time step $t$, the input gate $i_t$ is calculated as follows:
\begin{equation}
    i_t = \sigma(W_i \cdot [h_{t-1}, x_t] + b_i)\
\end{equation}
The input gate $i_t$ determines how much of the new traffic data should be contributed to the continuing congestion status after evaluating the existing traffic conditions, including the incoming packet rate, packet sizes, and network load. The forget gate $f_t$ describes how much information of the prior cell state should be maintained. It is in charge of controlling memory in the LSTM cell. The forget gate in the proposed model determines how important prior congestion information stored in the cell state is for the current congestion management decision. It is calculated as:
\begin{equation}
    f_t = \sigma(W_f \cdot [h_{t-1}, x_t] + b_f)\
\end{equation}
The candidate cell $C_{\tilde{t}}$ calculates the updated data that will be included in the cell state. It considers incoming data packet characteristics, including their sizes, types, and origins. It is calculated as:
\begin{equation}
    C_{\tilde{t}} = \tanh(W_c \cdot [h_{t-1}, x_t] + b_c)\
\end{equation}
Depending on the forget and input gates, the cell state $C_t$ integrates the previous cell state with the candidate cell state. It controls how information about congestion moves throughout the LSTM cell. It replicates how past congestion patterns interact with the current traffic situation to influence the choice of congestion control. It is calculated as:
\begin{equation}
    C_t = f_t \cdot C_{t-1} + i_t \cdot C_{\tilde{t}}\
\end{equation}
The essential information that the LSTM has learned from the sequence is contained in the hidden state $h_t$. The hidden state $h_t$ contains knowledge of congestion patterns across time in the proposed congestion control paradigm. The classification stage uses this hidden information to forecast congestion levels and inform congestion control measures. It is determined as:
\begin{equation}
    h_t = o_t \cdot \tanh(C_t)\
\end{equation}
The quantity of cell state that will be disclosed as the output is determined by the output gate $o_t$. The output gate $o_t$ evaluates the cell state, which represents the current congestion situation, and determines how useful this information is for making congestion management decisions. It is calculated as follows:
\begin{equation}
    o_t = \sigma(W_o \cdot [h_{t-1}, x_t] + b_o)\
\end{equation}
The Input Layer receives network data sequences that contain encoded category variables and normalized numerical characteristics. The LSTM Layers stack different LSTM layers to identify patterns and temporal relationships in the sequence data. Depending on the network traffic. In this work, we modify and adjust the number of LSTM units and layers to obtain better performance.
The output layer is a fully connected layer that generates congestion predictions by using a softmax activation function. The result of the softmax operation offers a probability distribution over different congestion levels. The predicted congestion level is determined by choosing the class with the highest likelihood, as shown in Figure \ref{fig:lstm-model}.

\subsection{Training}
We utilize the preprocessed data to train the LSTM network. The network develops the ability to anticipate congestion events based on past trends during training. We categorize congestion levels into low, medium, and high categories in the setting of IoT congestion control. To optimize the model, use the Adam optimizer and categorical cross-entropy.To improve the performance of the model, it was trained with a variety of hyperparameters, including the number of LSTM layers, units, and learning rates. Additionally, we use methods like early halting to avoid overfitting. The model is trained for 90 epochs. To make sure the model is learning efficiently, track training loss and validation statistics. After training, assess the effectiveness of the model on a test dataset.

\subsection{Congestion Prediction}
After training, use the trained LSTM model to estimate congestion levels in real-time by utilizing received network traffic data. To make judgments about congestion management automatically, create an adaptive control system that incorporates the LSTM predictions. This connection makes the congestion control system more adaptable, which improves the network's efficiency and user experience.

\section{Experiments}
We used a combination of actual network traffic traces and simulation device interactions to gather an extensive dataset tailored to IoT contexts for testing. It is gathered from IoT devices, such as sensors, actuators, and gateways, deployed in a setting of smart buildings.  The data included timestamps for congestion incidence, data rates, and packet arrival timings. The preprocessed dataset is separated into training 80\%, validation 10\%, and testing sets 10\%. The training set is utilized to train the LSTM model, the validation set assists in hyperparameter adjusting and prevents overfitting, and the testing set assesses the performance of the proposed model. The dataset was then organized into input sequences, each of which included the preceding 10 time steps as input features.
For the purpose of classifying congestion, an LSTM network architecture is created and set up. Because IoT data is temporal in nature, a stacked LSTM model with two layers is used for accurately capturing complex temporal correlations and patterns. In this work, each layer contains 64 LSTM cells.
The first layer acts as an extractor of features, while the second layer processes the extracted features to increase the accuracy of classification. In order to avoid overfitting, a dropout layer with a rate of 0.2 was placed between the LSTM layers. The output layer is a dense layer with a softmax activation function that generates the three congestion states. A learning rate of 0.001 is used to train the model by applying the Adam optimizer. The model is trained for 90 epochs with a batch size of 32. The model's weights and biases are optimized during training using backpropagation and gradient descent approaches. The parameters of the model are optimized using the categorical cross-entropy loss function.

\subsection{Evaluation Metrics}
Several performance measures are utilized to assess the success of the proposed approach.  The pace at which data is effectively sent via a communication channel during a certain time period is referred to as throughput. In this work, it is measured in bits per second (Kbps). It is calculated as:
\begin{equation}
  T = \frac{N}{RTT}  
\end{equation}
Where $N$ represents the number of bits transmitted and $RTT$ denotes the round-trip time. The amount of time it requires for data in a network to travel from source to destination is referred to as network delay $D$. It consists of several components that contribute to the total delay encountered by data packets as they travel across the network architecture. It is calculated as:
\begin{equation}
    D = \text{$P_D$} + \text{$Tr_D$} + \text{$Q_D$} + \text{$Pr_D$}
\end{equation}
Where $P_D$ represents the propagation delay, $Tr_D$ denotes the transmission delay, $Q_D$ shows queueing delay and $Pr_D$ illustrates the processing delay.

\subsection{Experiment Design}
Diverse network load situations, such as those with low, medium, and high traffic volumes, are used to perform the experiments. Performance data was recorded while each scenario was conducted for five minutes. The IoT dataset is supplied into the trained LSTM model for each scenario to forecast congestion. These predictions are then included in congestion control strategies. We gathered performance data and assessed the efficacy of each strategy.

\section{Results and Discussion}

The quantitative results demonstrated that the LSTM-enhanced congestion control technique consistently beat the baseline methods in terms of throughput, latency, and packet loss rate. Furthermore, the LSTM-enhanced method consistently produced greater user satisfaction ratings. The LSTM predictions achieved in this work for network congestion control in an IoT network are shown in Table \ref{results}. 
\begin{table*}[htbp]
\centering
\caption{LSTM prediction for congestion control}
\begin{tabular}{cccc}
\hline
Time (seconds) & Throughput (Kbps) & LSTM Congestion Prediction & Congestion Control Action \\
\hline
10 & 59 & 0.15 (Below threshold) & None \\
20 & 52 & 0.12 (Below threshold) & None \\
30 & 65 & 0.25 (Below threshold) & None \\
40 & 70 & 0.68 (Above threshold) & Traffic shaping \\
50 & 65 & 0.50 (Above threshold) & QoS adjustment \\
\hline
\end{tabular}
\label{results}
\end{table*}

\begin{figure}[t]
    \centering
    \includegraphics[width=0.9\linewidth]{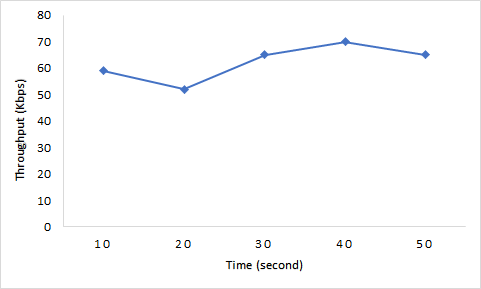}
    \caption{Achieved throughput using the proposed model}
    \label{fig:throughput}
\end{figure}

Device data is gathered by the IoT network at varying intervals. On the basis of the gathered data, the LSTM model is trained to forecast congestion. The likelihood of congestion is then predicted for each interval of time. Congestion is classified as substantial when it exceeds a preset threshold of (0.5). No congestion management measures are implemented when the prediction falls below the threshold (0.5). Congestion control measures are started when the forecast is greater than the threshold (0.5). Since there is a considerable possibility of congestion at 40 seconds, traffic shaping is used to regulate data rates. Although the projection is still higher than the threshold at 50 seconds, it is lower than the earlier prediction. As a result, a QoS adjustment is made to give certain traffic priority over others. Leveraging these predictions, the IoT network may manage congestion issues proactively, guaranteeing a better user experience and more effective use of network resources. The attained throughput in this work shown in Figure \ref{fig:throughput} is strongly related to the efficacy of managing network congestion to enable efficient data flow and a smooth user experience in the context of congestion control.

\subsection{Comparison}

\begin{figure}[t]
    \centering
    \includegraphics[width=0.9\linewidth]{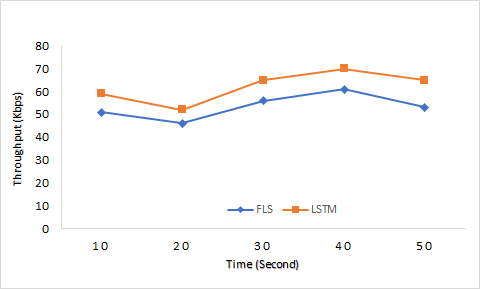}
    \caption{Achieved throughput using the proposed model and FLS scheme}
    \label{fig:Time-throughput}
\end{figure}

\begin{figure}[t]
    \centering
    \includegraphics[width=0.9\linewidth]{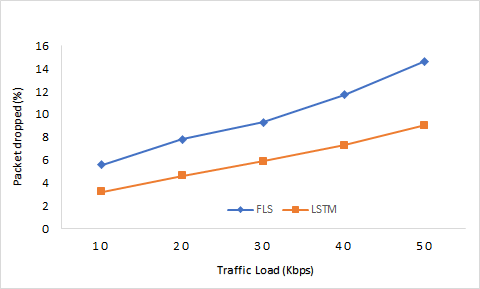}
    \caption{Traffic load Vs packet loss using the proposed model and FLS scheme}
    \label{fig:load-packet}
\end{figure}

The capacity of the LSTM model to accurately represent temporal patterns, together with its integration with control techniques, resulted in increased network efficiency and user experience. In order to validate the model performance, we compared the proposed work with \cite{aimtongkham2021enhanced}, which used a fuzzy logic system (FLS) for IoT Congestion Control. Their approach includes a multidimensional congestion predictor that detects congestion circumstances using a relative strength index and trend analysis, as well as an adaptable and congestion-aware backoff approach. We used the FLS approach and the proposed LSTM-based approach to obtain the throughput of the network, shown in Figure \ref{fig:Time-throughput}.
We additionally analyze the model using packet loss versus traffic load parameters using both the proposed model and the FLS scheme. We experience fewer packet drops than the FLS system, demonstrated in Figure \ref{fig:load-packet}.

\begin{figure}[t]
    \centering
    \includegraphics[width=0.9\linewidth]{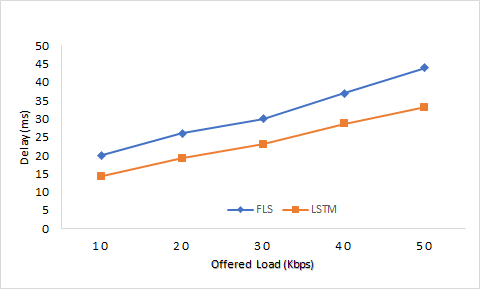}
    \caption{Offered load Vs network delay using the proposed model and FLS scheme}
    \label{fig:Delay-offerload}
\end{figure}

Lower network latency is frequently an important aim in network optimization since it immediately leads to enhanced user experience and more efficient data delivery. Because of the predictive nature of LSTM, fast responses and optimized resource allocation are possible, resulting in a reduced latency than the comparative FLS system, as shown in Figure \ref{fig:Delay-offerload}.

This accomplishment demonstrates the suggested model's effectiveness in preserving network stability and performance even during times of intense network congestion.

The suggested approach displayed improved adaptation to shifting network circumstances, resulting in faster data transport and fewer instances of congestion.
The displayed load reflects different amounts of load in Kbps, while the average delay indicates the delay in milliseconds.

\section{Conclusion}
This study illustrates the capability of LSTM networks to improve congestion control in IoT environments. It concentrated on how to better control the data flow in scenarios when several IoT devices are attempting to send and receive data. The proposed strategy for controlling congestion in IoT settings was evaluated using two primary approaches: theoretical analysis and practical experiments. We observed that the proposed model has an extraordinary capacity to reduce packet loss rates and maintain network performance even under varying traffic loads via rigorous testing and assessment. The suggested technique successfully improved network performance by utilizing LSTM predictions, highlighting its importance in guaranteeing stable IoT connection and raising satisfaction among users. In the future, we want to incorporate attention mechanisms or hybrid models, to further refine prediction accuracy.


\bibliographystyle{IEEEtran}
\bibliography{references}
\end{document}